\documentclass[twocolumn]{aa}
\usepackage{amssymb}

\usepackage{graphicx}
\usepackage{txfonts}
\usepackage{lipsum}
\usepackage{subcaption}         
\usepackage{xcolor}
\usepackage{lscape}             
\usepackage{placeins}           
                                
\usepackage[colorlinks=true, linkcolor=blue, urlcolor=blue, citecolor=blue]{hyperref}
\usepackage{array}

\usepackage{booktabs}

\def\ms{\hbox{m s$^{-1}$}}         
\def\gcm3{\hbox{g cm$^{-3}$}}       
\def\Msun{\hbox{$\mathrm{M}_{\odot}$}}             

\def\Mjup{\hbox{$\mathrm{M}_{\rm Jup}$}}

\newcommand{\gaia}{\emph{Gaia}\xspace}
\newcommand\gdr[1]{\gaia~DR#1}

\begin{document}


   \title{Love Is in The Air: CARMENES meets Gaia DR3 and its babies\thanks{Based on observations collected at Centro Astronómico Hispano en Andalucía (CAHA) at Calar Alto, proposal 26A-3.5-017 and proposal 26B-3.5-015, operated jointly by Junta de Andalucía and Consejo Superior de Investigaciones Científicas (IAA-CSIC).}}
   \subtitle{Confirmation of 5 Gaia DR3 substellar candidates with CARMENES}


%
%
%

   \author{E. Esparza-Borges\inst{1}\corrauth{Emma.Esparza-Borges@esa.int}        
        \and J. Lillo-Box\inst{2}\corrauth{Jorge.Lillo@cab.inta-csic.es}\fnmsep
        \and J. Sahlmann\inst{1}\email{}
        \and S. Turrado\inst{2}\email{}
        \and E. Delgado-Mena\inst{2}\email{}
        \and P.A. Miles-Páez\inst{2}
        \and O.~Balsalobre-Ruza\inst{3}
        \and C.~Cifuentes\inst{2}
        \and A. Masson\inst{4}
        \and I. Mendigutía\inst{2}
        \and I. Rebollido\inst{2}
        \and G. Roccetti\inst{1,5}
        \and C. Haukes\inst{2}
        }

   \institute{European Space Agency (ESA), European Space Astronomy Centre (ESAC), Camino Bajo del Castillo s/n, E-28692 Villanueva de la Cañada, Madrid, Spain
   \and Centro de Astrobiología (CAB), CSIC-INTA, Camino Bajo del Castillo s/n, 28692 Villanueva de la Cañada (Madrid), Spain
   \and Max-Planck-Institut f\"ur Astronomie, K\"onigstuhl 17, 69117 Heidelberg, Germany
   \and Universit\'e Paris-Saclay, Universit\'e Paris Cit\'e, CEA, CNRS, AIM, 91191, Gif-sur-Yvette, France
   \and INAF – Osservatorio Astrofisico di Arcetri, Largo E. Fermi 5, 50125 Firenze, Italy.}


   \date{Received November 30, 2026}

 
  \abstract
   {The fourth \textit{Gaia} data release (DR4) is expected to contain thousands of extrasolar and substellar candidate signals. Given the properties of the astrometric technique, these signals will cover orbital periods from hundreds to few thousand days. Confirming their masses and nature, as well as characterizing their properties and completing the architecture of those systems relies on ground-based radial velocity efforts.}
  {In preparation for Gaia DR4, we aim at evaluating the observational radial velocity effort required to confirm and improve the precision on the astrometric mass and orbital elements (including the eccentricity and argument of the periastron) and to look for additional inner lower-mass planetary or substellar companions in six \textit{Gaia} DR3 substellar companion candidates in temperate orbits (100-1000 days).} 
   {We use the CARMENES instrument to carry out a radial-velocity follow-up over a one-year time span using a simple long-term cadence of around 15-20 days.}
   {We confirm the substellar nature of five out of six candidates, discarding the sixth as an astrophysical false positive in the form of a near-equal brightness binary star. Four of them lay in the brown-dwarf desert, with masses spanning between $msin{i}=$ 13~$\rm M_{Jup}$ to $msin{i}=$ 55~$\rm M_{Jup}$. In one case we also identify a potential inner companion.
   }
   {Our observations show the clear benefit and need of radial velocity follow-up of astrometry-discovered candidates found in Gaia DR3 and more broadly those that will come with DR4, with particular emphasis on planetary signals, even more subject to false positives and worse parameter precision than the substellar objects studied in this work.  }


   \maketitle
   \nolinenumbers


\section{Introduction}


Long-period planets ($\rm P > 200~days$) have been elusive along the 30 years of exoplanet exploration. Their transit probability is negligible (only a few candidates have been detected by the Kepler mission) and they require very long-term monitoring with the radial velocity technique. The latter has only been achieved by GTO programs running blind search surveys along 20 years with, e.g., CORALIE \citep[e.g.,][]{Fontanet2025,Parc2026} or HIRES \citep[e.g.,][]{Blunt2019}. However, a blind search for very long-period giant planets (cold Jupiters) is inefficient and their detection only comes out as a byproduct of programs searching for shorter period planets and subsequently following-up long-term trends in the data. Direct imaging searches are relatively limited in their accessible target list (young and close-by stars) and the discoveries are sporadic and heavily dependent on the planet's orbital position.

Now, the European Gaia mission in its third data release \citep[DR3; ][]{GaiaCollaboration2016,GaiaCollaboration2023_A1} and its upcoming fourth data release (DR4) provide access, for the first time, to a homogeneous, all-sky, and population-scale sample of long-period giant planets and substellar companions through their astrometric signatures \citep{Perryman2014,Sozzetti2014,Vioque026}. Access to this population is fundamental for understanding the formation and dynamical evolution of planetary systems. When present, long-period gas giants and substellar companions often dominate the gravitational interaction with other planets, ultimately defining the architecture of the planetary system \citep{Chatterjee2008,Ikoma2025} and influencing the potential habitability of inner rocky planets \citep[e.g.,][]{Kane2024}. Statistical studies have revealed an asymmetrical relation between cold Jupiters and close-in super-Earths: the majority of systems with a long-period gas giant like host inner super-Earths, whereas only a minority of systems with close-in super-Earths host a detected cold Jupiter \citep{Zhu2018,Bryan2019}.

Gaia DR3 provided a list of more than 1800 stars with Gaia astrometric orbits compatible with long-period sub-stellar companions. Its revolutionary accuracy made possible to determine a large number of astrometric orbits with semi-major axes smaller than 1~AU for the first time \citep{Halbwachs2023}, finally opening the astrometric exoplanet science window \citep{SahlmannGomez25}. These companions also probe the so-called brown-dwarf desert, which represents an observed scarcity of brown-dwarf companions relative to both giant planets and stellar binaries at orbital separations within $\sim$3-5~AU \citep[e.g.,][]{Grether2006}. However, ground-based spectroscopic follow-up remains essential to confirm these candidates, constraining their masses and orbital parameters, or to rule out astrophysical false positives \citep{Marcussen2026}. 


In this work we show the benefits of astrometric guidance for RV searches of long-period sub-stellar companions, and the strengths of combining both techniques to better characterise masses and orbits. Here we used CARMENES to do a radial velocity follow-up of six Gaia DR3 substellar companion candidates. The substellar companions confirmed in this article lay in the underpopulated brown-dwarf region of the mass-period diagram, with masses spanning between 22-57~$\rm M_{Jup}$ and orbital periods between 226-436~days. 
We describe the observations and data reduction in Section~\ref{sec:obs} and the spectroscopic stellar characterization in Section~\ref{sec:Stellar_characterization_spectroscopy}. The methods and results are presented in Section~\ref{sec:analysisandresults}. Finally, we discuss and conclude our study in Section~\ref{sec:discussionandconclusions}. 







\section{Observations and data reduction}\label{sec:obs}

\subsection{Gaia DR3 and Target Sample}

Based on 34 months of observations and released on 13 June 2022, Gaia DR3 \citep{GaiaCollaboration2016,GaiaCollaboration2023_A1} delivered an astrometric census of about 1.8 billion targets, including $\sim$ 170\,000 compatible with two-body astrometrical solutions. Among them, more than 1800 long-period sub-stellar candidate companions were identified by fitting astrometric orbits with non-single star (NSS) models. These candidates are listed in the \texttt{gaiadr3.nss\_two\_body\_orbit} table, which also provides their Keplerian-orbit solutions. Complementary studies have also used machine learning approaches on Gaia DR3 data to identify the best potential candidates for subsequent follow-up \citep{SahlmannGomez25}.

In this study, we target substellar companion candidates in temperate orbits amenable for radial velocity follow-up with CARMENES spectrograph. Selected targets fulfill the following criteria: 
i) Presence of astrometric companions in the sub-stellar regime ($\rm <80~M_{Jup}$) detected by Gaia DR3;
ii) Main-sequence stars with $G\leq13.5$ (which turned out being F, G, or K -type stars);
iii) orbital period ranging from 100 to 1000 days; and
iv) visible in the Northern hemisphere (DEC~$>0^{\circ}$).

We list our target sample in Table~\ref{tab:gaia_targets}. Although initially we included Gaia\,DR3\,2081900940499099136 in the sample, we removed it from our target list after only a few spectra were taken because they showed significant line-broadening due to its high rotation velocity, preventing high-precision RV studies. A total of six targets were finally included in the full RV follow-up campaign presented in this work.

\begin{table*}[h] 
\caption{Properties of the selected Gaia DR3 systems from orbital solutions in \texttt{$gaiadr3.nss\_two\_body\_orbit$} table.} \label{tab:gaia_targets} 
\centering
\resizebox{\textwidth}{!}{
\begin{tabular}{lcccccccccc} 
\hline\hline
Gaia DR3 source ID & Name$^{(1)}$ & $G$ & SpT $^{(2)}$ & $P$ & $e$ & $\pi$ & $d_{\rm GSP-Phot}$ & $T_0$ & $i$ & $\omega^{(3)}$ \\
&  & (mag) & & (days) &  & (mas) & (pc) & (BJD) & (deg) & (deg) \\ 
\hline 
\object{Gaia DR3 1977546227497119872} & GDR3-1977 & $12.986$ & K5V & $224.11 \pm 0.86$ & $0.61 \pm 0.04$ & $8.48 \pm 0.03$  & $115.8_{-0.9}^{+0.8}$ & $2457392.0 \pm 2.3$ & $154 \pm 7$ & 10.88 \\[3pt]  
\object{Gaia DR3 2259327108945413888} & GDR3-2259 & $13.260$ & K5V & $365.76 \pm 4.72$ & $0.73 \pm 0.11$ & $5.20 \pm 0.02$  & $185.6_{-1.3}^{+1.1}$ & $2457213.9\pm15.1$ & $112 \pm 7$ & 339.57\\[3pt] 
\object{Gaia DR3 2280560705703031552} & HD 212620 & $8.705$ & F8V & $429.01 \pm 6.31$ & $0.58 \pm 0.14$ & $7.32 \pm 0.02$ & $135.5_{-0.4}^{+0.5}$ & $2457221.5\pm10.5$ & $124 \pm 5$ & 16.49\\[3pt] 
\object{Gaia DR3 3913728032959687424} & HD 99251 & $7.459$ & F5 & $328.90 \pm 3.09$ & $0.26 \pm 0.11$ & $14.75 \pm 0.07$  & $67.8_{-0.4}^{+0.7}$ & $2457485.7\pm12.9$ & $115 \pm 7$ & 150.61 \\[3pt] 
\object{Gaia DR3 4593569060397933952} & GDR3-4593 & $13.290$ & K5V & $433.75 \pm 3.44$ & $0.48 \pm 0.06$ & $5.97 \pm 0.02$ & $167.4_{-1.8}^{+3.2}$ & $2457351.4\pm9.6$ & $46 \pm 5$ & 154.44 \\[3pt] 
\object{Gaia DR3 4598289263814843136} & GDR3-4598 & $12.856$ & K7V & $435.81 \pm 4.30$ & $0.40 \pm 0.06$ & $6.15 \pm 0.02$ & $167.9_{-0.5}^{+2.7}$ & $2457544.2\pm11.2$ & $151 \pm 7$ & 350.42\\ \hline 
\end{tabular}
}
\tablefoot{$^{(1)}$ Name used in this article for simplicity and benefit of the reader. $^{(2)}$ Spectral Type from Simbad\footnote{\url{https://simbad.cds.unistra.fr/simbad/}}.
$^{(3)}$ Estimated from Gaia DR3 Thiele-Innes coefficients as described in \cite{Halbwachs2023}.}
\end{table*}

\subsection{CARMENES Observations}

We monitored the six Gaia DR3 targets with sub-stellar companion candidates through a Long-term Programme with CARMENES in semesters 2026A (Proposal ID: 26A-3.5-017) and 2026B (Proposal ID: 26B-3.5-015). CARMENES is a high-resolution spectrograph mounted on the 3.5 m telescope at the Calar Alto Observatory in Spain. It has two channels: the visual (VIS) and the near-infrared (NIR), that cover the 0.52–0.96 and 0.96–1.71 µm wavelength ranges, at spectral resolutions of $\mathcal{R}=94\,600$ and $\mathcal{R}=80\,400$, respectively \citep{Quirrenbach2014,Quirrenbach2018}.

The six targets were observed between January and September 2026, with between five and seven epochs per target. Gaia\,DR3\,1977546227497119872 (hereafter GDR3-1977) and Gaia\,DR3\,2259327108945413888 (GDR3-2259) were each observed five times with exposure times of 1800~s, reaching median VIS S/N values of 17 and 10, respectively. Gaia\,DR3\,2280560705703031552 (GDR3-2280) and Gaia\,DR3\,3913728032959687424 (GDR3-3913, HD\,99251) were observed six and five times, respectively, with shorter exposure times of 720--1460~s and 340-400~s, and reached substantially higher median VIS S/N values of 92 and 78. Gaia\,DR3\,4593569060397933952 (GDR3-4593) and Gaia\,DR3\,4598289263814843136 (GDR3-4598) were observed six and seven times, respectively, with 1800~s exposures, yielding median VIS S/N values of 12 and 14.







The CARMENES spectra were reduced using the standard \texttt{CARACAL} pipeline \citep{Zechmeister2014,Bauer2015}, which performs bias subtraction, flat-field correction, background subtraction, wavelength calibration, and extraction of the individual echelle spectral orders. Instrumental drifts were monitored through simultaneous Fabry-Pérot calibration exposures obtained with the second fibre (fibre B), and barycentric corrections were applied to the wavelength solution.

Stellar radial velocities (RVs) were subsequently extracted from the visible-channel (VIS) spectra using the \texttt{SHAQ} pipeline \citep{LilloBox2022,LilloBox2024}, based on the cross-correlation function (CCF) technique \citep{Baranne1996}. \texttt{SHAQ} computes the CCFs using spectral-type-dependent binary masks provided by the ESPRESSO pipeline \citep{Pepe2021}, with individual spectral lines weighted according to their depths. We used the binary masks corresponding to the spectral-type of each star according Table~\ref{tab:gaia_targets}. Spectral orders strongly affected by telluric contamination or displaying low signal-to-noise ratios were excluded from the analysis. The stellar RVs were determined from the centroid of a Gaussian profile fitted to the resulting CCFs, after correcting for instrumental drifts.

Additionally, several CCF diagnostics were extracted, including the full width at half maximum (FWHM), contrast, and bisector inverse span (BIS; \citealt{Lafarga2020}), providing complementary information to assess potential stellar activity and line-profile variations.

\section{Spectroscopic Stellar Characterization}\label{sec:Stellar_characterization_spectroscopy}
Individual stellar spectra were moved to the rest frame and combined with the SERVAL pipeline \citep{Zechmeister_2018} and converted from vacuum to air wavelengths using the conversion in \cite{Birch_Downs_1994}, and spectral orders were stitched together at the middle point of the overlapping wavelength region of each pair of orders. As a final step, the spectra were linearly resampled to a uniform $0.02\ \AA$ wavelength grid. In the case of GDR3-2259, one single-night spectrum was used due to SERVAL yielding unreasonably high calibrated fluxes for a large portion of spectral orders of this target.

Stellar parameters have been derived using the ionization balance method described in \cite{Sousa_2008}, which measures equivalent-widths of Fe\,I and Fe\,II atomic spectral lines with the ARES software \citep{Sousa_2007}, and calculates the Fe abundance from each line using the MOOG radiative transfer code \citep{Sneden_1973} and 1D LTE atmospheres from MARCS models \citep{Gustaffson_2008} interpolated between the nodes, using a least-squares solver to test combinations of atmospheric parameters $\left\{T_{\mathrm{eff}},\ \mathrm{\left[M/H\right]},\ \log g,\ \xi \right\}$ (where $\xi$ designates the microturbulence velocity) until iron line abundances reach ionization balance. Since ARES+MOOG measures absolute iron abundance $\mathrm{A}\left(\mathrm{Fe}\right)$, we convert the mean iron abundance measured  to relative solar $\mathrm{\left[Fe/H\right]}$ by subtracting $\mathrm{A}_\odot\left(\mathrm{Fe}\right)=7.47\ \mathrm{dex}$, and demand that it agrees with $\mathrm{\left[M/H\right]}$ from the model atmosphere to within $0.05\ \mathrm{dex}$, at which point we set $\mathrm{\left[M/H\right]}=\mathrm{\left[Fe/H\right]}$. Fe\,I lines that yield atmospheric abundances $2\sigma$ away from the mean abundance calculated from Fe\,I lines only are removed at each step of the least squares solver. According to \cite{Delgado_Mena_2017}, $\log g$ was additionally corrected from systematic errors dependent on the derived $T_{\mathrm{eff}}$, with en empirical relation based on trigonometric $\log g$. Additionally, for stars whose effective temperatures fell below $5200\ \mathrm{K}$ during the least-squares fit, the line list in \cite{Tsantaki_2013} was used.

Since spectroscopic $\log g$ is known to be unreliable \citep{Freckelton_2026}, we additionally calculated $\log g$ following the corrector-predictor method presented in \cite{Santos_2004}, which uses the targets \textit{Gaia}\,DR3 $G$ band brightness and parallax with the bolometric correction from \cite{Andrae_2018}, and  a mass estimation from the regression in \cite{Torres_2009}, applying the correction of \cite{Santos_2013}. This regression uses $T_{\mathrm{eff}}$, $\log g$ and $\mathrm{\left[M/H\right]}$ as input. This is further motivated by the low amount of Fe\,II lines available to constrain surface gravity with the EW-Ion balance method, leading to large errors (see Table \ref{tab:stellar_parameters}). The resulting $\log g_{\mathrm{trig}}$ can be finally input to the \cite{Torres_2009} formula to get a stellar mass estimation.

The errors of the magnitudes calculated from EW-Ion balance ($T_{\mathrm{eff}},\ \mathrm{\left[M/H\right]},\ \log g,\ \xi $) were obtained in the following way: first, following \cite{Gonzalez_Vanture_1998}, we calculated the deviations resulting from parameter perturbations of $100\ \mathrm{K}$, $0.1\ \mathrm{km/s}$ and $0.2\ \mathrm{dex}$ in $T_\mathrm{eff}$, $\xi$ and $\mathrm{\left[Fe/H\right]}$, respectively. Then, we added an epistemic error in quadrature to each parameter, as in section 3.1 of \cite{Sousa_2011}. This is done to account for the fact that the true parameters are unknown and that different methods yield different parameters, often outside of mutual error bars. Lastly, the error in $\log g_{\mathrm{trig}}$ and stellar mass was calculated by the bootstrap method, taking, for each magnitude involved in the \cite{Santos_2004} calculation, 1000 realizations of a gaussian distribution centered at the measured value and with standard deviation equal to the error. This includes a random mass draw from the \cite{Torres_2009} empirical relation itself, using the scatter of the regression as the standard deviation.

The target parameters calculated with the equivalent-width method are summarized in Table \ref{tab:stellar_parameters}. All of the target parameters are compatible with dwarf stars of subsolar metallicity in the main sequence. This was the case for every target except for GDR3-3913. Of the 298 iron lines selected from HARPS spectra in the \cite{Sousa_2008} line list, 137 Fe\,I and 14 Fe\,II lines fall within the wavelength range of the reduced CARMENES spectra. For the \cite{Tsantaki_2013} line list, 76 Fe\,I and 6 Fe\,II lines fall within the range. Due to this scarcity of available Fe\,II lines, we did not apply sigma-clipping to the Fe\,II lines. However, for GDR3-4593 and GDR3-4598 convergence was not reached in $\log g$ using this approach, so we decided to eliminate Fe\,II lines from the fit one by one, starting from the line with the highest absolute deviation in abundance from the mean, until convergence was reached. In the case of GDR3-3913, convergence could not be reached in $\xi$ due to some Fe\,I lines with extreme equivalent-width values, which had to be removed to reach convergence.

\begin{table*}[!h] 
\caption{Stellar atmospheric parameters and iron abundances derived from the equivalent-width analysis.} 
\label{tab:stellar_parameters} 
\centering \resizebox{\textwidth}{!}{
\begin{tabular}{lllllllllllll}
\hline\hline 
\textit{Gaia} DR3
& SNR & $T_{\mathrm{eff}}$ & $\mathrm{[Fe/H]}$ & $\log g$      & $\log g_{\mathrm{trig}}$ & $\xi$          & $N_{\ion{Fe}{i}}$ & $N_{\ion{Fe}{ii}}$ & $\sigma_{\ion{Fe}{i}}$ & $\sigma_{\ion{Fe}{ii}}$ & Mass            & Age             \\
                  source ID  &     & (K)                & (dex)             & (dex)         & (dex)                    & (km\,s$^{-1}$) &                   &                    & (dex)                  & (dex)                   & ($M_\odot$)       & (Gyr)           \\ 
\hline 
GDR3-1977 & 49  & $4380\pm210$       & $-0.287\pm0.081$  & $4.58\pm0.59$ & $4.79\pm0.16$            & 0.0+1.6        & 69                & 4                  & 0.22                   & 0.33                    & $0.610\pm0.061$ & $4.49 \pm 4.04$ \\
GDR3-2259 & 25  & $4860\pm150$       & $-0.171\pm0.066$  & $4.88\pm0.52$ & $4.807\pm0.091$          & $1.40\pm0.33$  & 75                & 6                  & 0.17                   & 0.55                    & $0.713\pm0.050$ & $3.95 \pm 3.84$ \\
HD~99251 & 275 & $6910\pm230$       & $-0.12\pm0.12$    & $4.22\pm0.19$ & $4.350\pm0.073$          & $3.57\pm1.56$  & 65                & 8                  & 0.19                   & 0.16                    & $1.348\pm0.093$ & $0.90 \pm 0.74$ \\
GDR3-4593 & 49  & $4470\pm130$       & $-0.241\pm0.057$  & $4.49\pm0.39$ & $4.677\pm0.097$          & $0.83\pm0.31$  & 66                & 4                  & 0.15                   & 0.30                    & $0.639\pm0.047$ & $4.43 \pm 4.04$ \\
GDR3-4598 & 55  & $4630\pm90$        & $-0.323\pm0.045$  & $4.80\pm0.16$ & $4.653\pm0.066$          & 0.0+1.2        & 59                & 3                  & 0.072                  & 0.057                   & $0.666\pm0.043$ & $4.47 \pm 3.98$ \\ \hline
\end{tabular} } 
\tablefoot{ The name of each object follows the definition in Table~\ref{tab:gaia_targets}. $T_{\mathrm{eff}}$ is the stellar effective temperature; $\log g$ is the surface gravity obtained through the EW-Ion balance method after applying the correction of \cite{Delgado_Mena_2017}; $\log g_{\mathrm{trig}}$ is the surface gravity calculated from \textit{Gaia} DR3 parallaxes following the methodology of \cite{Santos_2004} with the empirical mass estimation of \cite{Torres_2009}; $\xi$ is the microturbulence velocity; and $\mathrm{[Fe/H]}$ is the stellar iron abundance relative to the Sun. $N_{\ion{Fe}{i}}$ and $N_{\ion{Fe}{ii}}$ are the numbers of Fe\,I and Fe\,II lines that ARES was able to measure for the analysis, respectively, while $\sigma_{\rm FeI}$ and $\sigma_{\rm FeII}$ are the standard deviation of iron abundance coming from Fe\,I and Fe\,II lines only, respectively. Mass is the stellar mass from the empirical relation of \cite{Torres_2009}. Age is the stellar age calculated by PARAM 1.3 \citep{dasilva06}.}
\end{table*}

The microturbulence velocity $\xi$ could not be constrained for GDR3-4598 and GDR3-1977, due to the combination of a high $\mathrm{A}\left(\mathrm{Fe}\right)$ scatter due to low signal-to-noise (see Table \ref{tab:stellar_parameters}) and the fact that, at such low values of $\xi$, thermal broadening dominates and $\xi$ barely changes the abundance versus reduced equivalent-width slope over the line list. Due to this, the microturbulence velocity errors for these targets are only $2\sigma$ upper-limits.

The two calculations of $\log g$ discussed above agree within mutual error bars, but the trigonometric surface gravity of \cite{Santos_2004} is more precise, so we used the latter throughout this work. We acknowledge that such a calculation could be affected by bad \textit{Gaia} DR3 photometry or parallax from possible stellar companions; however, the suspected substellar nature of the companions proposed in this paper remains in agreement with the choice of trigonometric $\log g$. Furthermore, we note that most of the targets in this paper turned out to be late K dwarfs, which are at the edge of the validity range of the empirical mass calculation of \cite{Torres_2009}.

Finally we derived the stellar masses and ages with the PARAM 1.3 tool\footnote{\url{http://stev.oapd.inaf.it/cgi-bin/param_1.3}} which makes use of the PARSEC theoretical isochrones \citep{bressan12} and a Bayesian estimation method \citep{dasilva06}, together with the spectroscopic $T_{\mathrm{eff}}$, [Fe/H], the magnitude $V$ and the parallax from Gaia DR3. The values obtained for the K dwarfs are quite uncertain but expected, especially considering the large errors in $T_{\mathrm{eff}}$.

\section{Analysis and Results}\label{sec:analysisandresults}

\subsection{Methods}

Given the limited number of CARMENES spectra available for each target, we adopted an informed Bayesian approach to model the RV time series, taking advantage of the orbital constraints provided by \gaia DR3 astrometry. As discussed by \citet{SahlmannGomez25}, spectroscopic follow-up is suitable to validate \gaia astrometric orbital solutions and constrain the nature of the companions. We modeled the RVs using a single-Keplerian model parametrized by the orbital period ($P$), the time of periastron passage ($T_{\rm p}$), the eccentricity ($e$), the argument of periastron ($\omega$), and the RV semi-amplitude ($K$), together with an instrumental offset ($\delta$) and jitter ($\sigma$). To incorporate the astrometric information, we adopted Gaussian priors on $P$, $T_{\rm p}$, $e$, and $\omega$, centered on the corresponding \gaia DR3 orbital solutions, with standard deviations derived from their reported uncertainties (Table~\ref{table:priors}). In contrast, uniform priors were assigned to the RV semi-amplitude and instrumental offset, allowing these parameters to be constrained by the spectroscopic observations. The posterior probability distributions were explored using a Markov chain Monte Carlo (MCMC) approach implemented through the \texttt{emcee} ensemble sampler \citep{ForemanMackey2013}, following the general methodology described in \citet{LilloBox2024}. This approach allowed us to exploit the astrometric constraints to reduce the orbital parameter space while independently assessing the RV amplitude and its uncertainty, providing additional constraints on the nature and minimum mass of the companions. In cases were the solution shows clearly large residuals or significant differences of the orbital elements or physical parameters with respect to the astrometric solution, models with both a second Keplerian (labelled as "2k") and a linear trend (labelled as "LT") are also tested. In such cases, given the small number of RV measurements, the second Keplerian is kept with a circular orbit, thus adding only three additional parameters to the model.

We used the MCMC chains to obtain the marginalised posterior distributions for each parameter. The median and 68.7\% confidence interval are reported in Table~\ref{table:results} as our final reported parameters. 
To assess the statistical significance of the Keplerian signals, we estimated the Bayesian evidence ($\mathcal{Z}$) of the different models from the posterior distributions sampled with \texttt{emcee}. For this purpose, we used the \texttt{perrakis} implementation \citep{Diaz2016} of the importance-sampling method developed by \citet{perrakis14}, following the approach adopted in previous studies \citep[e.g., ][]{LilloBox2021}. This method estimates the marginal likelihood by constructing an importance-sampling distribution from the product of the marginal posterior distributions of the model parameters, thus allowing the Bayesian evidence to be computed directly from the converged MCMC chains without requiring additional posterior sampling. For each model, we used samples from the post-burn-in chains to estimate the logarithmic Bayesian evidence ($\ln \mathcal{Z}$) and its numerical distribution. We then compared the competing models through the difference in their log-evidences, $\Delta\ln\mathcal{Z}=\ln\mathcal{Z}_i-\ln\mathcal{Z}_j$, equivalent to the logarithm of the Bayes factor. Positive values favor model $i$ over model $j$, with $\Delta\ln\mathcal{Z}>6$ indicating strong statistical evidence in favor of the former model.

The posterior distributions and derived parameters for each of the studied systems are presented in Table~\ref{table:results}.

\begin{table*}
\setlength{\extrarowheight}{3pt}
\small
\caption{Prior distributions assumed for the orbital and stellar parameters of the five Gaia DR3 targets.}
\label{table:priors}
\centering
\resizebox{\textwidth}{!}{ 
\begin{tabular}{lccccc}
\hline\hline
\noalign{\smallskip}
 Parameter & {GDR3-1977} & {GDR3-2259} & {HD 99251 } & {GDR3-4593} & {GDR3-4598} \\
\hline
\noalign{\smallskip}
$P_b$ [days] & $\mathcal{G}(224.12, 0.86)$ & $\mathcal{G}(365.76, 4.72)$ & $\mathcal{G}(328.90, 3.09)$ & $\mathcal{G}(433.75, 3.44)$ & $\mathcal{G}(435.81, 4.30)$ \\
$T_{\rm 0,b} - 2400000$ [days] & $\mathcal{G}(57403.98, 4.99)$ & $\mathcal{G}(57203.80, 17.03)$ & $\mathcal{G}(57451.61, 19.94)$ & $\mathcal{G}(57421.46, 18.39)$ & $\mathcal{G}(57608.18, 26.21)$ \\
$K_b$ [m/s] & $\mathcal{U}(0.0, 10000.0)$ & $\mathcal{U}(0.0, 10000.0)$ & $\mathcal{U}(0.0, 10000.0)$ & $\mathcal{U}(0.0,3000.0)$ & $\mathcal{U}(0.0, 10000.0)$ \\
$e_b$ & $\mathcal{T}(0.61, 0.04, 0, 1)$ & $\mathcal{T}(0.73, 0.11, 0, 1)$ & $\mathcal{T}(0.26, 0.11, 0, 1)$ & $\mathcal{T}(0.48, 0.06, 0, 1)$ & $\mathcal{T}(0.40, 0.06, 0, 1)$ \\
$\omega_b$ [deg.] & $\mathcal{G}(10.88, 15.75)$ & $\mathcal{G}(159.57, 25.65)$ & $\mathcal{G}(150.61, 19.74)$ & $\mathcal{G}(154.44,9.68)$ & $\mathcal{G}(350.42, 20.61)$ \\
\noalign{\smallskip}
\noalign{\smallskip}
$\delta_{\rm CARMENES}$ [m/s] & $\mathcal{U}(-150.0, 150.0)$ & $\mathcal{U}(-150.0, 150.0)$ & $\mathcal{U}(-150.0, 150.0)$ & $\mathcal{U}(-150.0, 150.0)$ & $\mathcal{U}(-150.0, 150.0)$ \\
$\sigma_{\rm CARMENES}$ [m/s] & $\mathcal{B}(1.9, 3.0)$ & $\mathcal{B}(1.9, 3.0)$ & $\mathcal{B}(1.9, 3.0)$ & $\mathcal{B}(1.9, 3.0)$ & $\mathcal{B}(1.9, 3.0)$ \\
Slope [m~s$^{-1}$~day$^{-1}$] & $\mathcal{U}(-500.0, 500.0)$ & \dots & $\mathcal{U}(-500.0, 500.0)$ & \dots & \dots \\
\hline
\end{tabular}}
\tablefoot{
$\mathcal{G}$ denotes Gaussian, $\mathcal{U}$ Uniform, $\mathcal{T}$ Truncated Gaussian, and $\mathcal{B}$ Beta distribution. The name of each object follows the definition in Table~\ref{tab:gaia_targets}.}
\end{table*}

\begin{table*}
\setlength{\extrarowheight}{3pt}
\caption{Posterior distributions of the orbital and physical parameters derived for the five Gaia DR3 substellar companion candidates.}
\label{table:results}
\centering
\resizebox{\textwidth}{!}{
\begin{tabular}{lcccccc}
\hline\hline
\noalign{\smallskip}
\bf Gaia DR3 source ID$^\ast$ & \bf {GDR3-1977} & \bf {GDR3-2259} & \multicolumn{2}{c}{\bf {HD 99251}} & \bf {GDR3-4593} & \bf {GDR3-4598} \\
\hline
\noalign{\smallskip}
Model$^\dagger$ & 1kQT & 1k & 1kLT & 2k & 1k & 1k \\
\bf \textit{Companion Orbital Parameters} & & &  &  & & \\
$P_b$ [days] & $226.07^{+0.31}_{-0.30}$ & $360.3^{+1.4}_{-1.4}$ & $333.8^{+1.6}_{-1.6}$ & $337.6^{+1.7}_{-1.6}$ & $426.8^{+1.7}_{-1.8}$ & $435.7^{+3.4}_{-3.3}$ \\
$T_{\rm 0,b} - 2400000$ [days] & $57407.7^{+4.8}_{-4.8}$ & $57197^{+16}_{-15}$ & $57470^{+17}_{-17}$ & $57489^{+16}_{-17}$ & $57400^{+16}_{-16}$ & $57607^{+23}_{-22}$ \\
$K_b$ [m/s] & $760^{+131}_{-96}$ & $3000^{+2040}_{-920}$ & $7680^{+580}_{-450}$ & $4700^{+1310}_{-820}$ & $1374^{+54}_{-41}$ & $1680^{+1020}_{-730}$ \\
$e_b$ & $0.610^{+0.046}_{-0.045}$ & $0.756^{+0.085}_{-0.086}$ & $0.038^{+0.017}_{-0.016}$ & $0.212^{+0.086}_{-0.113}$ & $0.450^{+0.020}_{-0.017}$ & $0.411^{+0.060}_{-0.062}$ \\
$\omega_b$ [deg.] & $0^{+11}_{-11}$ & $160.4^{+3.7}_{-2.8}$ & $157^{+19}_{-21}$ & $141^{+19}_{-20}$ & $148.7^{+7.6}_{-7.3}$ & $348^{+15}_{-17}$ \\
\noalign{\smallskip}
\bf \textit{Instrumental Parameters} & & & & & & \\
$\delta_{\rm CARMENES}$ [km/s] & $-39.066^{+0.085}_{-0.086}$ & $-16.562^{+0.047}_{-0.079}$ & $-20.51^{+0.45}_{-0.47}$ & $-14.58^{+0.43}_{-0.74}$ & $-13.606^{+0.065}_{-0.059}$ & $-10.54^{+0.54}_{-0.45}$ \\
$\sigma_{\rm CARMENES}$ [m/s] & $7400^{+8600}_{-5100}$ & $6.4^{+7.5}_{-4.4}$ & $5.0^{+6.6}_{-3.5}$ & $5200^{+6800}_{-3500}$ & $6500^{+7700}_{-4600}$ & $9.8^{+11.2}_{-6.9}$ \\
Slope [m~s$^{-1}$~day$^{-1}$] & $26.0^{+3.0}_{-2.9}$ & \dots & $106.5^{+8.9}_{-6.5}$ & \dots & \dots & \dots \\
Quadratic [$m^2~s^{-2}$~day$^{-2}$] & $-0.127^{+0.015}_{-0.016}$ & \dots & \dots & \dots & \dots & \dots\\
\noalign{\smallskip}
\bf \textit{Companion Physical Parameters} & & & & & & \\
$m_{b}\sin{i_b}$ [$M_{\rm Jup}$] & $13.0^{+1.9}_{-1.6}$ & $55^{+22}_{-12}$ & $320^{+28}_{-25}$ & $127^{+36}_{-24}$ & $33.8^{+1.9}_{-1.9}$ & $44^{+26}_{-19}$ \\
$a_b$ [AU] & $0.616^{+0.020}_{-0.021}$ & $0.885^{+0.020}_{-0.021}$ & $1.04^{+0.024}_{-0.025}$ & $1.048^{+0.024}_{-0.025}$ & $0.956^{+0.023}_{-0.024}$ & $0.982^{+0.021}_{-0.022}$ \\
$a_b/R_{\star}$ & $189.1^{+8.4}_{-8.3}$ & $250.3^{+9.0}_{-8.7}$ & $203.3^{+5.1}_{-5.2}$ & $204.9^{+5.1}_{-5.2}$ & $293^{+11}_{-11}$ & $302^{+11}_{-11}$ \\
$F_{\rm inc,b}$ [$F_{\rm inc,\oplus}$] & $0.426^{+0.098}_{-0.082}$ & $0.368^{+0.056}_{-0.049}$ & $2.29^{+0.35}_{-0.31}$ & $2.25^{+0.34}_{-0.30}$ & $0.192^{+0.028}_{-0.025}$ & $0.209^{+0.023}_{-0.021}$ \\
$T_{\rm eq,b}$ [K] & $205^{+11}_{-11}$ & $197.9^{+7.1}_{-7.0}$ & $312^{+11}_{-11}$ & $311^{+11}_{-11}$ & $168.2^{+5.9}_{-5.8}$ & $171.8^{+4.6}_{-4.5}$ \\
\noalign{\smallskip}
\bf \textit{Additional Companion Orbital Parameters} & & & & & & \\
$P_c$ [days] & \dots & \dots & \dots & $90.5^{+5.8}_{-6.0}$ &  & \dots \\
$T_{\rm 0,c} - 2400000$ [days] & \dots & \dots & \dots & $61122.08^{+0.68}_{-0.63}$ &  & \dots \\
$K_c$ [m/s] & \dots & \dots & \dots & $780^{+130}_{-140}$ &  & \dots \\
\noalign{\smallskip}
\bf \textit{Additional Companion Physical Parameters} & & & & & & \\
$m_{c}\sin{i_c}$ [$M_{\rm Jup}$] & \dots & \dots & \dots & $32.0^{+5.0}_{-5.3}$ &  & \dots \\
$a_c$ [AU] & \dots & \dots & \dots &  $0.435^{+0.021}_{-0.021}$ &  & \dots \\
$a_c/R_{\star}$ & \dots & \dots & \dots & $85.1^{+4.1}_{-4.1}$ &  & \dots \\
$F_{\rm inc,c}$ [$F_{\rm inc,\oplus}$] & \dots & \dots & \dots & $13.1^{+2.3}_{-2.0}$ &  & \dots \\
$T_{\rm eq,c}$ [K] & \dots & \dots & \dots & $483^{+20}_{-19}$ &  & \dots \\ \hline
$\Delta\ln{\mathcal{Z}}^{\ddagger}$ & +26.8 & \dots & +7.0 & +18.9 & \dots & \dots \\
\hline

\end{tabular}}
\tablefoot{$^\ast$ The name of each object follows the definition in Table~\ref{tab:gaia_targets}. $^\dagger$ Scenario finally adopted for each target, including only 1-Keplerian (1k) as in the astrometric solution, adding a linear slope (1kLT), adding a quadratic term (1kQT) or adding a second Keplerian (2k).
$^\ddagger$ Difference in the log-Bayesian evidence of the corresponding model (stated in the first row of the table) and the nominal astrometric solution of just 1-keplerian signal. (1k).
}
\end{table*}

\subsection{Confirmed substellar companions}

\begin{figure*}[!htbp]
    \centering
    \includegraphics[width=\linewidth]{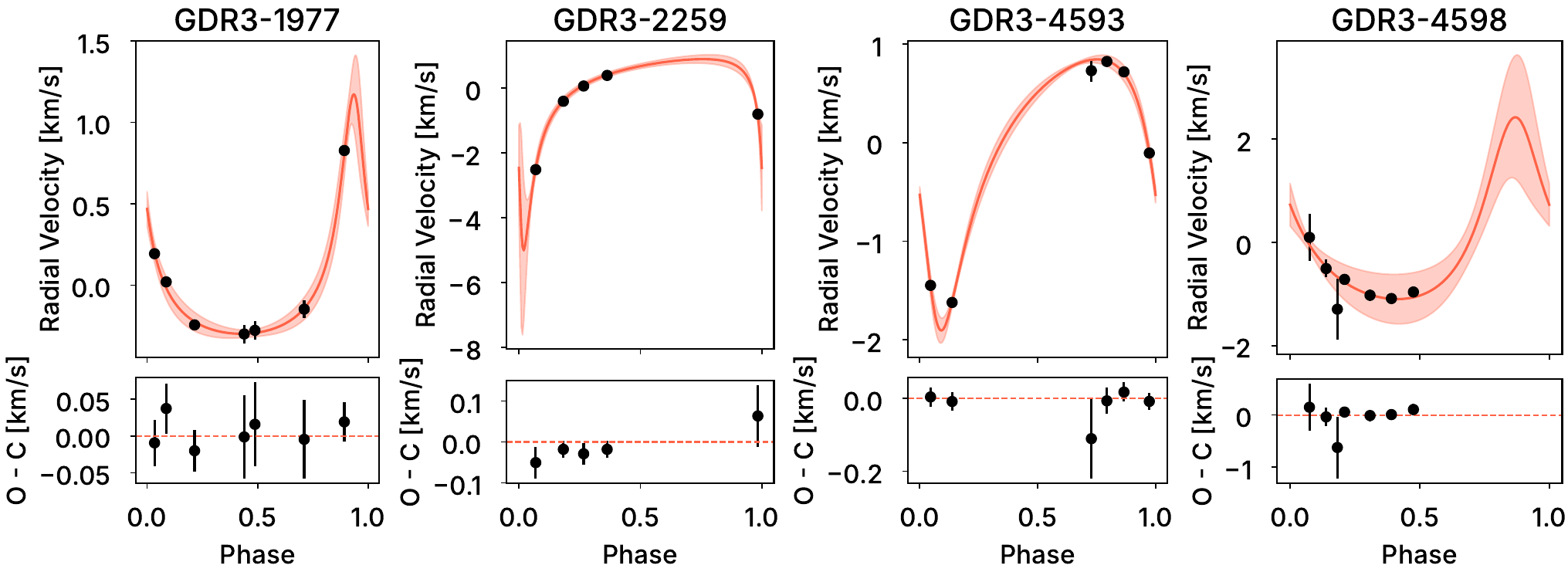}
    \includegraphics[width=\linewidth]{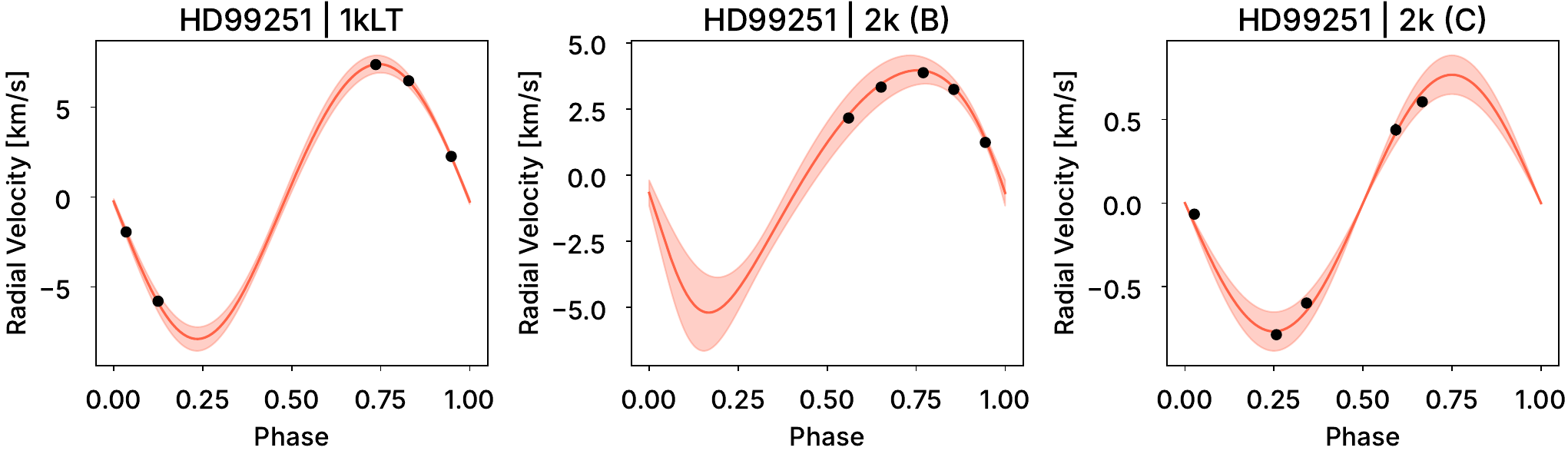}
    \caption{CARMENES RV measurements for all confirmed \gaia DR3 substellar companions (black points). Solid lines show the best-fit Keplerian model and shaded regions show posterior distributions within $1\sigma$. Top row: Substellar candidates confirmed with no additional companions. Middle row: The case of HD99251, with the 1-component model including a linear trend (left panel) and the two individual components: HD99251\,b (middle panel, corresponding to the astrometric solution found by \citealt{SahlmannGomez25}) and HD99251\,c (right panel, corresponding to the new potential detection). Bottom row: The same as in the middle row but for the target GDR3-4593.}
    \label{fig:RV_mosaic}
\end{figure*}

From our list of six Gaia DR3 systems, we confirmed the substellar nature of the companion for five targets and identified one false positive (described in Section~\ref{sec:FP}). We show the RV curves for all confirmed candidates in Figure~\ref{fig:RV_mosaic} and describe each target below.

\paragraph{GDR3-1977.-} For GDR3-1977, the 1-Keplerian model (labelled "1k") produces a relatively large jitter of 34 m/s, suggesting the need for additional components in the data. We first try the addition of a linear trend ("1kLT") that still keeps the jitter value at the same level. Instead, the addition of a quatdratic trend ("1kQT") decreases the jitter posterior to a truncated gaussian at zero. The goodness of this model including the Keplerian and the quadratic trend versus the previous simpler models is clear from the Bayesian evidence, which provides a value of +26.8 when compared to the simple 1-Keplerian model and +25.8 when compared to the model adding the linear trend. To test whether the data suports the detection of a sencond Keplerian, we tried a fourth model including only two Keplerians ("2k"). This more complex model, however, is not supported by the data against the quadratic trend model, with $\ln{\mathcal{Z}_{\rm 2k-1kQT}}=-22$). Hence, we take the "1kQT" model as the one being more supported by the current dataset. The results for this model on the orbital elements show a good agreement with the Gaussian priors. The semi-amplitude of the radial velocity provides $K=760^{+131}_{-96}$~\ms, corresponding to $m_b\sin{i_b}=13.0^{+1.9}_{-1.6}$ and, assuming the inclination from the astrometric fit (i.e., $i_b=154\pm7.2$~deg), we obtain $m_b=30.1\pm9.0$~\Mjup. 

\paragraph{GDR3-2259.-} GDR3-2259 presents a highly eccentric ($e\sim0.76$) substellar companion as for the astrometric solutions, that is reflected in the CARMENES RVs. The RV modeling confirms this eccentric orbit although still lacks some good coverage at the periastron passage. This lack of data implies a large uncertainty in the inferred semi-amplitude and hence mass of the companion, which we report as $K=3000^{+2040}_{-920}$~\ms{} and $m_{b}\sin{i_b}=55^{+22}_{-12}$~\Mjup. Assuming the orbital inclination from Gaia DR3 -- $i_b=112.0\pm7.2$~deg -- we obtain $m_b=60\pm19$~\Mjup. No evidence for additional companions or linear trends are found in the current RV dataset. 

\paragraph{HD~99251 (GDR3-3913).-} The simple 1-Keplerian modelling of the RV for HD~99251 yields a jitter of 43 \ms, thus suggesting the presence of additional effects missing in the model. We  repeat the analysis using a linear trend. This results in a large decrease in the scatter of the residual RVs and a decrease of the jitter parameter below the RV uncertainties. Hence, we find this model clearly preferred, as also shown by the differential Bayesian evidence against the model without slope ($\Delta\ln\mathcal{Z}=+7$). The corresponding posterior of the slope provides $106.5^{+8.9}_{-6.5}$~\ms${\rm day}^{-1}$. Interestingly, the resulting orbital configuration provides a very low and well-constrained eccentricity of $0.038^{+0.017}_{-0.016}$, compared to the astrometric-inferred value of $0.26\pm0.11$. This solution produces a large RV semi-amplitude of $K=7680^{+580}_{-450}$~\ms, which for this F9V corresponds to a stellar M-dwarf companion with a mass of $0.305^{+0.027}_{-0.024}$~\Msun. This companion mass is significantly larger than the astrometry-only estimate under the dark companion assumption, but it is compatible with the companion mass range given in the \texttt{gaiadr3.binary\_masses} table \citep[][their Table 4]{SahlmannGomez25}. Because we find evidence for a third body in this system and the \gdr{3} astrometric orbit imposed a two-body model, the mass discrepancy can possibly be resolved by one companion being luminous, by a bias originating in the \gdr{3} modelling as a two-body system, or both.


\paragraph{GDR3-4593.-} The 1-Keplerian model for this system based on the astrometric solution provides a very good agreement with the RV data when accounting for the 180$^{\circ}$ degeneracy in the conversion of the Thiele-Innes elemented into the Campbell elements regarding the argumento of the periastron. In this case, the RV data clearly pinpoints the argument of the periastron to $148.7^{+7.6}_{-7.3}$ deg. From this model, the resulting minimum mass corresponds to $33.8\pm 1.9$~\Mjup. By taking into account the derived inclination from the astrometric solution, we obtain an absolute mass of $47.4\pm4.9$~\Mjup.

\paragraph{GDR3-4598.-}  For GDR3-4598, the RV analysis shows an excellent agreement of the RV data with the astrometric solution, providing a RV semi-amplitude of $1680^{+1020}_{-730}$~\ms, corresponding to a minimum mass of $44^{+26}_{-19}$~\Mjup. Assuming the orbital inclination from the astrometric solution ($i=151.0\pm7.5$~deg), the absolute mass of the companion translates into $91\pm53$~\Mjup. In this case, the high orbital inclination and the poor orbital coverage missing the periastron passage, prevents us from confirming the substellar nature of the companion, leaving opened the possibility that it is a low-mass late-type M-dwarf.

\subsection{False positive}\label{sec:FP}
\begin{figure}[h!]
    \centering
    \includegraphics[width=\linewidth]{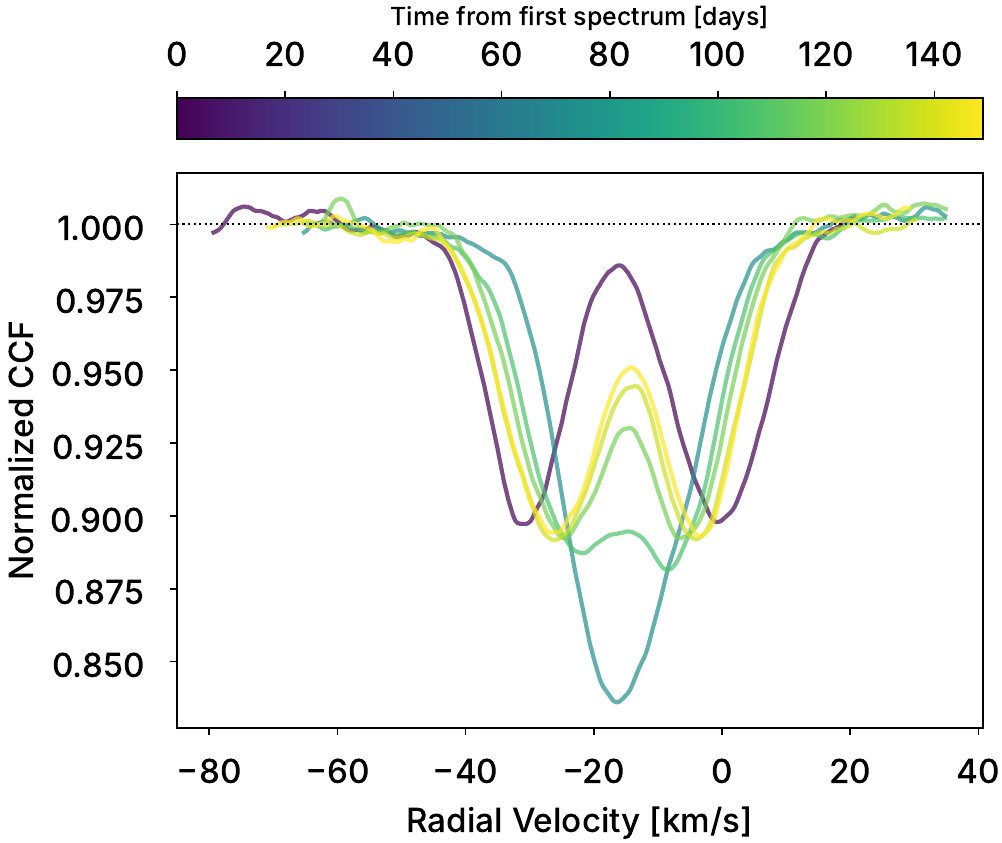}
    \caption{Cross-correlation functions (CCFs) from HD~212620 spectra observed between 22 April 2026 to 18 September 2026. Double peaks with same contrast reveal a false positive in the substellar companion of this Gaia DR3 source, actually being a spectroscopic binary star of type SB2 with both stars having approximately the same spectral type.}
    \label{fig:FP_CCF}
\end{figure}

The target HD~212620 (Gaia\,DR3\,2280560705703031552) represents a canonical example of the potential false positive configurations that astrometric observations can produce \citep[e.g.,][]{Marcussen2023}. The astrometric solution from Gaia DR3 produced a substellar candidate at $429.01 \pm 6.31$ days orbital period with a predicted dark-companion mass of 76.1 $\rm M_{Jup}$ \citep{SahlmannGomez25}. Our high-resolution spectroscopic measurements showed a clear double peak in the CCF on the first spectrum we obtained on 22 April 2026 (Figure~\ref{fig:FP_CCF}). The subsequent spectrum, obtained on 05 July 2026, however, displayed a single, deeper, CCF, showing a clear displacement of both components in a timespan of few months. We kept observing this target (with four additional spectra in the subsequent two months) to confirm the false positive nature of the candidate. Critically, both peaks have the same contrast, showing that both components have nearly equal spectroscopic features, thus being of similar spectral type. We thus confirm that the sub-stellar candidate identified in HD~212620 is actually a spectroscopic binary star of type SB2 with both stars having approximately the same spectral type. 

\section{Discussion and conclusions}\label{sec:discussionandconclusions}

We performed a radial-velocity follow-up of six \gaia~DR3 substellar companion candidates in temperate orbits (100 -- 1000 days) using CARMENES observations over a one-year baseline with a sparse cadence of 5 to 7 epochs per target.

\begin{figure*}[!htbp]
    \centering
    \includegraphics[width=\linewidth]{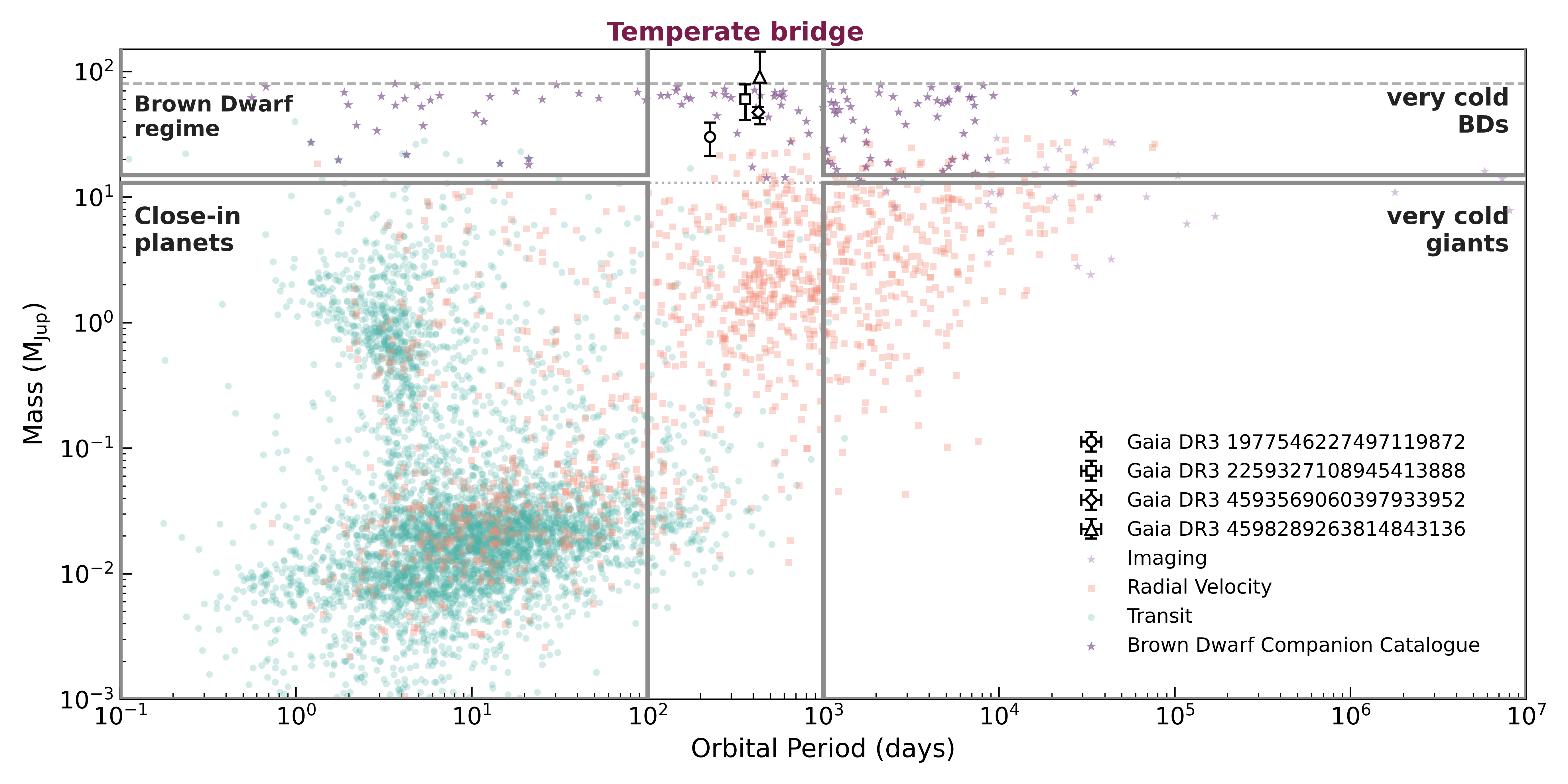}
    \caption{Confirmed substellar companions in the mass-period diagram, together with confirmed exoplanets through different detection methods from the NASA Exoplanet Archive \citep[][\url{https://exoplanetarchive.ipac.caltech.edu/}]{Christiansen2025} and the Brown Dwarf Companion Catalogue from \cite{Stevenson2023} (\url{https://doi.org/10.21954/ou.rd.24156393}).}
    \label{fig:MassPeriod}
\end{figure*}

\begin{figure*}[!htbp]
    \centering
    \includegraphics[width=0.8\linewidth]{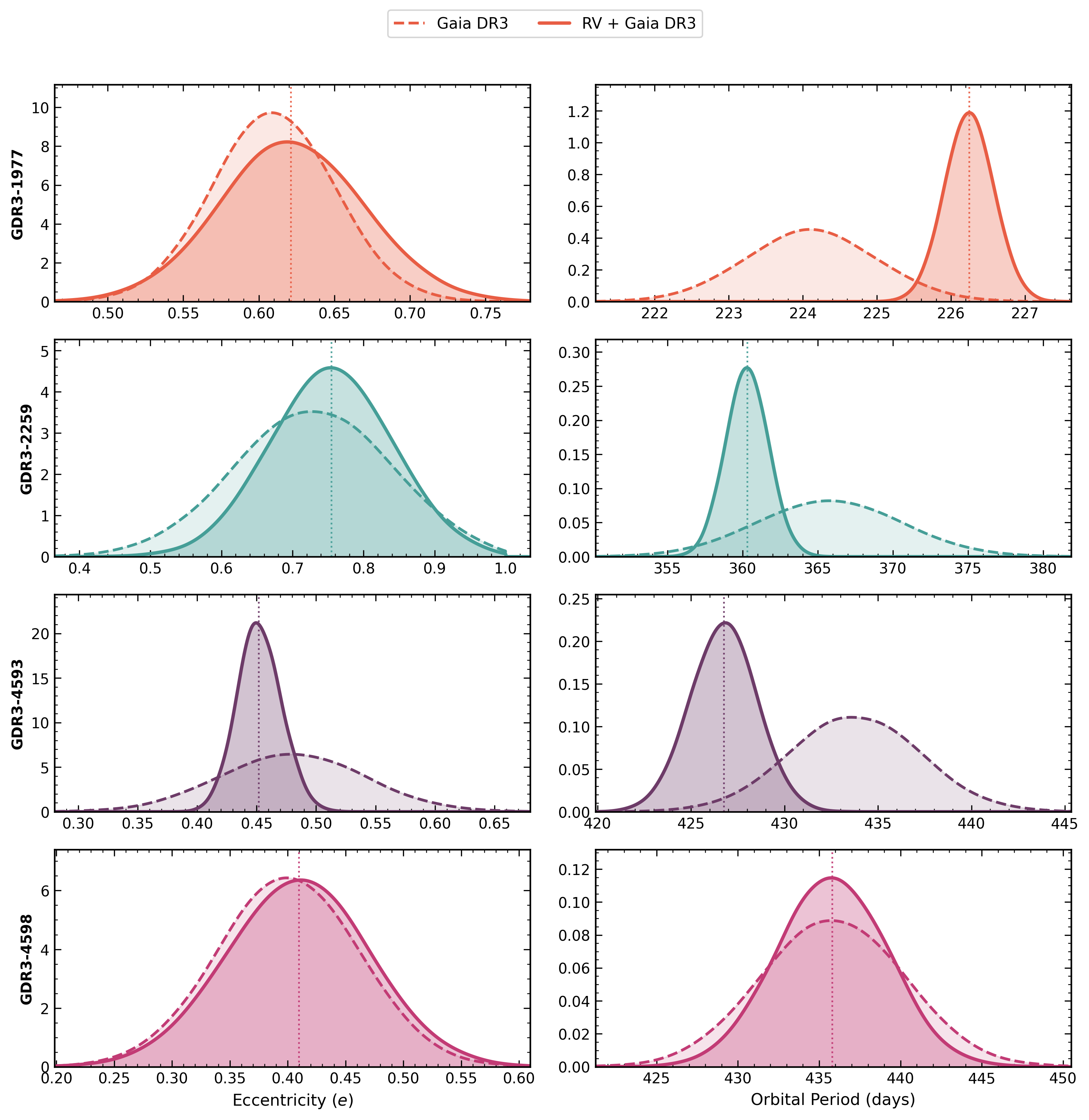}
    \caption{Comparison of \gaia DR3 astrometric priors from orbital solutions in \texttt{$gaiadr3.nss\_two\_body\_orbit$} table and RV posterior distributions. 
    }
    \label{fig:Hists}
\end{figure*}

We confirmed the substellar companions orbiting GDR3-1977 ($msin{i} = 13.0^{+1.9}_{-1.6} M_{\rm Jup}$), GDR3-2259 ($msin{i} = 55^{+22}_{-12} M_{\rm Jup}$), GDR3-4593 ($msin{i} = 33.8^{+1.9}_{-1.9} M_{\rm Jup}$) and GDR3-4598 ($msin{i} = 44^{+26}_{-19} M_{\rm Jup}$). These four objects are within the brown dwarf regime and reside in the brown dwarf desert region in the mass-period diagram (Figure~\ref{fig:MassPeriod}). Beyond confirming their substellar nature, combining RV observations with \gaia astrometry allows for refining the orbital parameters with, for example, a significant reduction in the orbital period uncertainty of GDR3-2259 from 1.29\% to 0.39\% and a general improvement in the accuracy in the orbital period of all confirmed substellar companions (Figure~\ref{fig:Hists}). 

Using the estimated semi-major axes of the companions and distances of their hosts, their characteristic angular separations range from approximately 2 to 6 mas, with the orbital separations reaching up to about 9 mas near apoastron. Although these angular separations make direct imaging and atmospheric characterization challenging, their dynamically determined masses, together with the moderately subsolar metallicities of their host stars, make them interesting objects for testing substellar evolutionary models. This is particularly relevant in the subsolar metallicity regime, where objects with independently determined masses and ages remain scarce. Future improvements in age estimates for the host stars will help constrain the effective temperatures, radii, and surface gravities of these brown dwarfs with evolutionary models.


We found that RV measurements in HD~99251 are compatible with two scenarios: i) a two-body Keplerian solution in which the mass of the companion is significantly larger than the astrometry-only estimate and ii) a three-body Keplerian solution with an additional $\rm \sim 32~M_{Jup}$ companion in a  $\rm \sim 90~days$ period orbit. We find a higher Bayesian evidence ($\Delta\ln\mathcal{Z}=+18.9$) for the scenario with an additional inner companion, showing that astrometric-guided RV campaigns can be sensitive to multi-companion architectures below the astrometric detection threshold. Further RV follow-up observations are needed to disentangle these two scenarios and confirm or rule out the presence of the inner companion. If confirmed, the inner companion would be a very interesting object for atmospheric characterization as it occupies the driest region of the brown-dwarf desert. 


We discarded HD~212620 (Gaia DR3 2280560705703031552) as a false positive, in which small astrometric orbits were produced by a spectroscopic binary with both stars having near-equal spectral types. This result shows how false positive cases can be rapidly identified with a small number of RV observations, highlighting the need of RV follow-up for \gaia substellar and planetary candidate signals. Although in the low-number statistics regime, identifying one false positive out of six randomly selected candidates ($\sim 16$\%) is among the expectations from previous works, for instance \cite{Lammers2026} obtaining a 13\% false-positive rate from their DR4 simulations.

With \gaia DR4 expected to discover tens of thousands of exoplanet and substellar candidates \citep{Lammers2026}, our results demonstrate both the need and benefit of RV follow-up. Even a modest RV campaign (5-7 epochs) is sufficient to rapidly identify false-positives, detect inner companions and improving the precision in orbital parameters. In this context, RV follow-up through upcoming exoplanet-devoted facilities with long-term dedicated guarantee time programs  \citep[e.g., EXOTICA,][]{LilloBox2026_MAESM} will be critical to confirm exoplanet candidates when \gaia DR4 pushes astrometric detections into the planetary-mass regime.






\begin{acknowledgements}
This work has made use of data from the European Space Agency (ESA) mission {\it Gaia} (\url{https://www.cosmos.esa.int/gaia}), processed by the {\it Gaia} Data Processing and Analysis Consortium (DPAC, \url{https://www.cosmos.esa.int/web/gaia/dpac/consortium}). Funding for the DPAC has been provided by national institutions, in particular the institutions participating in the {\it Gaia} Multilateral Agreement. 
E. E-B. and G. R. acknowledge support through the European Space Agency (ESA) Research Fellowship Programme in Space Science.
J.L.-B., E.D.-M. and C.H. are funded by the Spanish Ministry of Science and Universities (MICIU/AEI/10.13039/501100011033) through grant PID2023-150468NB-I00.
A. M. acknowledges funding by the European Union (ERC, Exo-RARE, 101265472).
E.D.-M. \& S.T. acknowledge the support by the Ram\'on y Cajal contract RyC2022-035854-I funded by MICIU/AEI/10.13039/501100011033 and by ESF+ and by the PIE project 20245AT026 funded by CSIC.
P.A.M.P. acknowledges the use of grant RYC2021-031173-I, funded by MCIN/AEI/10.13039/501100011033 and the European Union NextGenerationEU/PRTR, as well as project PID2025-169716NB-C32, funded by the Spanish Ministerio de Ciencia, Innovación y Universidades.
I.M. is funded by grant PID2022-138366NA-I00, by
the Spanish Ministry of Science and Innovation/State Agency of Research
MCIN/AEI/10.13039/501100011033 and by the European Union.
\end{acknowledgements}

%

\bibliographystyle{aa} 
\bibliography{Biblio} 







   
  



\begin{appendix}
\onecolumn
\section{Radial velocity measurements}

Table~\ref{tab:rv_measurements} shows the RV measurements for all confirmed substellar companions. These measurements will be publicly available online.

\begin{table}[h!]
\caption{\label{tab:rv_measurements} CARMENES radial velocity measurements for the five confirmed Gaia DR3 substellar companions.}
\centering
\begin{tabular*}{\textwidth}{@{\extracolsep{\fill}} l c c @{}}
\hline\hline
\bf Target name & \bf BJD & \bf RV \\
 & (days) & ($\mathrm{km\,s^{-1}}$) \\
\hline
\bf GDR3-1977 & 2461185.5962022 & $-39.26 \pm 0.08$ \\
 & 2461226.6209096 & $-37.38 \pm 0.03$ \\
 & 2461258.5893286 & $-37.67 \pm 0.04$ \\
 & 2461270.5067669 & $-37.73 \pm 0.05$ \\
 & 2461299.5303632 & $-38.01 \pm 0.04$ \\
\hline
\bf GDR3-2259 & 2461154.6336222 & $-17.37 \pm 0.08$ \\
 & 2461185.4994280 & $-19.08 \pm 0.04$ \\
 & 2461226.4895471 & $-16.97 \pm 0.02$ \\
 & 2461256.5605166 & $-16.50 \pm 0.03$ \\
 & 2461291.4235228 & $-16.17 \pm 0.02$ \\
\hline
\bf HD~99251 & 2461053.6948280 & $-13.07 \pm 0.02$ \\
 & 2461084.5606017 & $-10.72 \pm 0.04$ \\
 & 2461124.4691625 & $-10.72 \pm 0.02$ \\
 & 2461153.4523847 & $-11.87 \pm 0.01$ \\
 & 2461183.3647601 & $-12.57 \pm 0.01$ \\
\hline
\bf GDR3-4593 & 2461124.6197507 & $-12.87593 \pm 0.10902$ \\
 & 2461152.6461627 & $-12.78238 \pm 0.03620$ \\
 & 2461183.6288091 & $-12.88767 \pm 0.02648$ \\
 & 2461229.4903256 & $-13.71104 \pm 0.02326$ \\
 & 2461260.4131546 & $-15.05272 \pm 0.02658$ \\
 & 2461299.4003337 & $-15.22598 \pm 0.02544$ \\
\hline
\bf GDR3-4598 & 2461124.5983276 & $-105.31 \pm 0.45$ \\
 & 2461152.6199555 & $-105.91 \pm 0.17$ \\
 & 2461171.5366317 & $-106.70 \pm 0.59$ \\
 & 2461183.5903170 & $-106.12 \pm 0.08$ \\
 & 2461226.4405799 & $-106.43 \pm 0.11$ \\
 & 2461262.4001698 & $-106.49 \pm 0.11$ \\
 & 2461299.3749675 & $-106.37 \pm 0.11$ \\
\hline
\end{tabular*}
\tablefoot{Target names follow the convention defined in Table~\ref{tab:gaia_targets}.}
\end{table}

\end{appendix}





%

\end{document}